\documentclass{article}

\usepackage{arxiv}

\usepackage[utf8]{inputenc} 
\usepackage[T1]{fontenc}    
\usepackage{hyperref}       
\usepackage{url}            
\usepackage{booktabs}       
\usepackage{amsfonts}       
\usepackage{nicefrac}       
\usepackage{microtype}      
\usepackage{lipsum}
\usepackage{graphicx}
\usepackage{amsthm}
\usepackage{amsmath}
\usepackage{algorithm}  
\usepackage{algorithmic}

\newtheorem{theorem}{Theorem}

\usepackage{bm}

\title{Low-dimensional Manifold Constrained Disentanglement Network for Metal Artifact Reduction}

\author{
Chuang Niu,
\ Wenxiang Cong,
\ Fenglei Fan,
\ Hongming Shan,
\ Mengzhou Li,
\ Jimin Liang,
\ Ge Wang
}

\begin{document}
\maketitle

\begin{abstract}

Deep neural network based methods have achieved promising results for CT metal artifact reduction (MAR), most of which use many synthesized paired images for training. As synthesized metal artifacts in CT images may not accurately reflect the clinical counterparts, an artifact disentanglement network (ADN) was proposed with unpaired clinical images directly, producing promising results on clinical datasets. However, without sufficient supervision, it is difficult for ADN to recover structural details of artifact-affected CT images based on adversarial losses only. To overcome these problems, here we propose a low-dimensional manifold (LDM) constrained disentanglement network (DN), leveraging the image characteristics that the patch manifold is generally low-dimensional. Specifically, we design an LDM-DN learning algorithm to empower the disentanglement network through optimizing the synergistic network loss functions while constraining the recovered images to be on a low-dimensional patch manifold. Moreover, learning from both paired and unpaired data, an efficient hybrid optimization scheme is proposed to further improve the MAR performance on clinical datasets. Extensive experiments demonstrate that the proposed LDM-DN approach can consistently improve the MAR performance in  paired and/or unpaired learning settings, outperforming competing methods on synthesized and clinical datasets.

\end{abstract}


\section{Introduction}

Metal objects in a patient, such as dental fillings, artificial hips, spine implants, and surgical clips, will significantly degrade the quality of computed tomography (CT) images. The main reason for such metal artifacts is that the metal objects in the field of view strongly attenuate x-rays or even completely block them so that reconstructed images from the compromised/incomplete data are corrupted in various ways, which are usually observed as bright or dark streaks. As a result, the metal artifacts significantly affect medical image analysis and subsequent clinical treatment. Particularly, the metal artifacts degrade the counters of the tumor and organs at risk, raising great challenges in determining a radio-therapeutic plan \cite{Gian2017,Maerz}.

Over the past decades, extensive research efforts \cite{Gjesteby2016} have been devoted to CT metal artifact reduction (MAR), leading to various of MAR methods.
Traditionally, the projection domain methods \cite{Kalender, nmar} focus on projection data inside a metal trace, and replace them with estimated data. Then, the artifact-reduced image can be reconstructed from the refined projection data using a reconstruction algorithm, such as filtered backprojection (FBP). However, the projection domain methods tend to produce secondary artifacts as it is difficult for the estimated projection values to perfectly match the ground truth. In practice, the original projection data and the corresponding reconstruction algorithm are not publicly accessible. To apply the projection based methods in the absence of original sinogram data, researchers such as reported in \cite{Bal} proposed a post-processing scheme that generates the sinogram through forward projection of a CT image first and then applies the projection based method on the reprojected sinogram. However, the second-round projection and reconstruction may introduce extra errors. 

To overcome the limitations of the projection domain methods, researchers worked extensively to reduce metal artifacts directly in the image domain \cite{Hamid,karimi}. With deep learning techniques \cite{deepimaging}, data-driven MAR methods were recently developed  based on deep neural networks \cite{Huang2018MetalAR,Wang2018,Gjesteby_2019,zhang2018,dudonet,adn}, demonstrating superiority over the traditional algorithms for metal artifact reduction. However, most existing deep learning based methods are fully-supervised, requiring a large number of paired training images, i.e., the artifact-affected image and the co-registered artifact-free image. In clinical scenarios, it is infeasible to acquire a large number of such paired images. Therefore, the prerequisite of these method is to simulate artifact-affected images by inserting metal objects into artifact-free images to obtain paired data. However, simulated images cannot reflect all real conditions due to the complex physical mechanism of metal artifacts and many technical factors of the imaging system, degrading the performance of the fully-supervised models. To avoid synthesized data, the recently proposed ADN \cite{adn} only uses clinical unpaired metal-affected and metal-free CT images to train a disentanglement network on adversarial losses, giving promising results on clinical datasets, outperforming the fully-supervised methods trained on the synthesized data. However, without accurate supervision, the proposed ADN method is far from being perfect, and cannot preserve structural details in many challenging cases.

In this study, we improve the MAR performance on clinical datasets from two aspects. First, we formulate the MAR as the artifact disentanglement while at the same time leveraging the low-dimensional patch manifold of image patches to help recover structural details. Specifically, we train a disentanglement network with ADN losses and simultaneously constrain a reconstructed artifact-free image to have a low-dimensional patch manifold. The idea is inspired by the low-dimensional manifold model (LDMM) \cite{LDMM} for image processing and CT image reconstruction \cite{cong2019}. However, how to apply the iterative LDMM algorithm to train the disentanglement network is not trivial. To this end, we carefully design an LDM-DN algorithm for simultaneously optimizing objective functions of the disentanglement network and the LDM constraint.
Second, we improve the MAR performance of the disentanglement network by integrating both unpaired and paired supervision. Specifically, the unpaired supervision is the same as that used in ADN, where unpaired images come from artifact-free and artifact-affected groups. The paired supervision relies on synthesized paired images to train the model in a pixel-to-pixel manner. Although the synthesized data cannot perfectly simulate the clinical scenarios, they still provide helpful information for recovering artifact-free images from artifact-affected ones. Finally, we design a hybrid training scheme to combine both the paired and paired supervision for further improving MAR performance on clinical datasets.

The rest of this paper is organized as follows. In the next section, we review the related work. In section \ref{sec_method}, we describe the proposed method, including the problem formulation, the construction of a patch set, the dimension of a patch manifold, the corresponding optimization algorithm, and the hybrid training scheme. In section \ref{sec_experiment}, we evaluate the proposed LDM-DN algorithm for MAR on synthesized and clinical datasets.
Extensive experiments show that our proposed method consistently outperforms the state-of-the-art ADN model and other competing methods.
Finally, we conclude the paper in section \ref{sec_conclusion}.


\section{Related work}

\subsection{Metal Artifact Reduction}
CT metal artifact reduction methods can be classified into three categories, including projection domain methods, image domain methods and dual domain methods.

Projection based methods aim to correct projections for MAR. Some methods of this type \cite{park2016, jiang2000, meyer2010} directly corrects corrupted data by modeling the underlying physical process, such as beam hardening and scattering. However, the results of these methods are not satisfactory when high-atom number metals are presented. Thus, a more sophisticated way is to treat metal-affected data within the metal traces as unreliable and replaced them with the surrogates estimated by reliable ones. Linear interpolation (LI) \cite{Kalender} is a basic and simple method for estimating metal corrupted data. The LI method is prone to generate new artifacts and distort structures due to mismatched values and geometry between the linearly interpolated data and the unaffected data. To address this problem, the prior information is employed to generate a more accurate surrogate in various ways \cite{Bal, nmar, wj2013}. Among these methods, the state-of-the-art normalized MAR (NMAR) method \cite{nmar} is widely used due to its simplicity and accuracy. NAMR introduces a prior image of tissue classification  for normalizing the projection data before the LI is operated. With the normalized projection data, the data mismatch caused by LI can be effectively reduced for better results than that of the generic LI method. However, the performance of NMAR largely depends on the accurate tissue classification. In practice, the tissue classification is not always accurate so that NMAR also tends to produce secondary artifacts. Recently, deep neural networks \cite{Bernhard2017,liao2019,Ghani2020} were applied for projection correction and achieved promising results. However, such learning based methods require a large number of paired projection data.

The image domain based methods directly reduce metal artifacts based on CT image post-processing. Conventionally, some methods \cite{Hamid,karimi} leverage image processing techniques to estimate and remove the streak artifacts from original artifact-affected images, but these hand-crafted methods have a limited performance. From a data-driven perspective, deep learning methods for MAR have superiority over traditional approaches. For example, RL-ARCNN \cite{Huang2018MetalAR} is a convolutional neural network with residual learning for MAR, achieving better results than the plain CNN \cite{vgg}. cGANMAR \cite{Wang2018} regards MAR as an image-to-image transformation, and adapts the Pix2pix \cite{Isola_2017_CVPR} model to improve the MAR performance. 

To benefit from both projection and image domains, various dual domain based methods were also proposed. DestreakNet \cite{Gjesteby_2019} takes a corrected image by the state-of-the-art NMAR method and a detail mapping derived from the original image as the inputs to a dual-stream network, giving better results than what achievable in a single domain. In the CNN-MAR method \cite{zhang2018}, the CNN first takes the original image and corrected images by BHC \cite{Verburg2012CTMA} and LI \cite{Kalender} as the inputs and then produces a CNN output image, which is used to generate a prior image. Then, the projection data of the prior image is used to correct the original projection data, and the final image is reconstructed with FBP. DuDoNet \cite{dudonet} introduces an end-to-end dual domain network to simultaneously correct sinogram data and CT images.

All above deep learning methods for MAR require a large number of synthesized paired project datasets and/or CT images for training. A recent study \cite{adn} has shown that the models \cite{zhang2018, Wang2018} trained on the synthesized data cannot generalize the well on the clinical datasets. Then, ADN was designed and tested on clinical unpaired data, achieving promising results. However, without a strong supervision, ADN can hardly recover structural details in challenging cases.

In this study, we introduce a novel image prior, i.e., low-dimensional manifold, and different levels of supervision to train the disentangle network for improving the MAR performance on clinical datasets. Our proposed LDM prior guided disentanglement framework and synergistic supervision scheme have the potential to empower image domain and dual domain based methods as further detailed below.

\subsection{Low-dimensional manifold}

The patch set of natural images has been proved coming from a low-dimensional manifold \cite{Lee2003,carlsson2008,peyre2008,peyre2009}. Based on this low dimensionality of the patch manifold, LDMM first computes the dimension of the patch manifold based on the differential geometry and then uses the dimension to regularize an image recovery problem, including image impainting, super-resolution, and denoising. Based on LDMM, LDMNet \cite{ldmnet} proposes to regularize the combination of input data and output features within a low-dimensional manifold in the context of the classification task, showing a competitive performance over popular regularizers such as low-rank and DropOut. Recently, Cong et al \cite{cong2019} proposed to use LDMM in regularizing the CT image reconstruction, demonstrating that the LDMM has a strong ability to recover detailed structures in CT images. Inspired by the recent results with LDMM, here we propose an LDM constrained disentanglement network with both paired and unpaired supervision for improving the MAR performance on clinical datasets.

\section{Method}
\label{sec_method}

\subsection{Problem formulation}
\label{sec_problem}

Before the problem formulation, let us introduce the general neural network based method in the image domain. In the supervised learning mode, we have the paired data $\{x_i, x_i^{gt}\}_{i=1}^N$ available, where each artifact-affected image $x_i$ , has a corresponding artifact-free image $x_i^{gt}$, and $N$ is the number of paired images. Then, the deep neural network based model can be trained on this dataset with the loss function:
\begin{equation}
    \label{eq_sup}
    \mathcal{L}_{sup}(\theta) = \frac{1}{N} \sum_{i=1}^N \ell(f_{\theta}(x_i), x_i^{gt})
\end{equation}
where $\ell$ is a loss function, such as the $L1$-distance function, and $f_{\theta}(x_i)$ represents the predicted image of $x_i$ by the neural network with a parameter vector $\theta$. In practice, a large number of paired data are synthesized for training the model, as the clinical datasets only contain unpaired images.

To improve the MAR performance on clinical datasets, ADN adapts the generative adversarial learning based disentanglement network for MAR, only requiring an unpaired dataset, $\{(x_i, y_j)\}, i=1, \cdots, N_1, j=1, \cdots, N_2$, where $y_i$ represents an artifact-free image. The ADN consists of several encoders and decoders, which are trained with several loss functions, including two adversarial losses, a reconstruction loss, a cycle-consistent loss and an artifact-consistent loss. For simplicity, we denote the ADN loss functions as:
\begin{equation}
    \label{eq_adn}
    \mathcal{L}_{adn}(\theta) = \frac{1}{N_1 N_2} \sum_{i=1}^{N_1} \sum_{j=1}^{N_2} \ell_{adn}(f_{\theta}(x_i), y_j)
\end{equation}
where $l_{adn}$ represents the combination of all loss functions of ADN.

In this work, we introduce a general image property of known as LDM to improve the MAR performance on clinical datasets. Specifically, we assume that a patch set of artifact-free images samples a low-dimensional manifold. 
Therefore, we formulate the MAR problem as follows:
\begin{equation}
    \label{eq_problem}
    \min_{\theta, \mathcal{M}} \mathcal{L}(\theta) + dim(\mathcal{M(P_{\theta})})
\end{equation}
where $P_{\theta}$ denotes the patch set of artifact-free or artifact-corrected images, $\mathcal{M}$ is a smooth manifold isometrically embedded in the patch space, and $\mathcal{L}(\theta)$ can be any network loss functions, such as $\mathcal{L}_{sup}$ for paired learning or $\mathcal{L}_{adn}$ for unpaired learning.

To solve the above optimization problem, we need to specify the reconstruction of a patch set, the computation of a patch manifold, and the learning algorithm for simultaneously optimizing the network functions and the dimensionality of a patch manifold. In the following sections, we will describe each of them.

\begin{figure}
    \centering
    \includegraphics[width=0.6\textwidth]{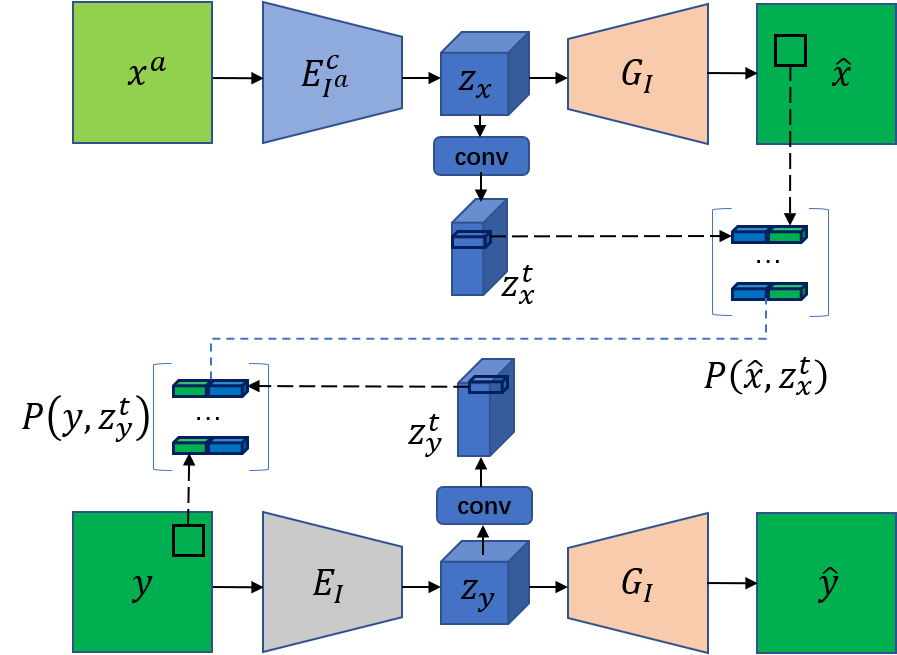}
    \caption{Diagram of patch set construction.}
    \label{fig:patch}
\end{figure}

\subsection{Construction of a patch set}
\label{sec_patch}
In this work, we adapt the state-of-the-art disentangle network in our proposed LDMM-based optimization framework under different levels of supervision.
For such a disentanglement network, we leverage its two branches to construct a patch set. As shown in Fig. \ref{fig:patch}, one is the artifact-corrected branch that maps artifact-affected images to artifact-corrected images, and the other is the artifact-free branch that maps artifact-free images to themselves.
Considering the spatial correspondence between the input/output image and its convolutional feature maps, we concatenate each feature vector along the spatial axes and its corresponding image patch to represent a patch. For the artifact-corrected branch, we take patches from the artifact-corrected images, denoted as $\{P_i(\hat{x})\}$. For the artifact-free branch, we take patches from the original images, denoted as $\{P_j(y)\}$. As we assume that the patch set of the images without artifacts samples a low-dimensional manifold, the final patch set is the concatenation of these two patch sets, denoted by $P_{\theta} = P(\hat{x}, z_x^t) \cup P(y, z_y^t)$. 

In our implementation, the input image size is $H\times W$ and the step size for down-sampling the encoder features is $s$, then the patch size is $s \times s$, the shape of $z_x^t$ or $z_y^t$ is $s^2 \times \frac{H}{s} \times \frac{W}{s}$, and each patch $P_{\theta}^i \in R^{2s^2}$.

\subsection{Dimension of patch manifold}
\label{sec_dimension}
In this work, we adopt the definition introduced in LDMM \cite{LDMM} for computing the dimension of a patch manifold. Specifically, we have the following theorem.

\begin{theorem}
    Let $M$ be a smooth submanifold isometrically embedded in $R^d$. For any patch $\bm{p}=(p_i)_{i=1}^d \in \mathcal{M}$,
    \begin{equation}
          dim(\mathcal{M})=\sum_{j=1}^d||\nabla_M\alpha_j(\bm{p})||^2 
    \end{equation}
\end{theorem}
where $\alpha_i(\bm{p})=p_i$ is the coordinate function, $\nabla_M\alpha_j(\bm{p})$ denotes the gradient of the function $\alpha_i$ on $\mathcal{M}$. More details on the definition of $\alpha_i$ on $\mathcal{M}$ can be found in \cite{LDMM}.
In our implementation, $\bm{p} = P_{\theta}^i \in \bm{R}_d, d=2s^2$, where the patch is parameterized by the neural network parameter vector $\theta$, as introduced in Section \ref{sec_patch}.

\subsection{Optimization}
\label{sec:opti}
According to the the construction of a patch set and the definition of a patch manifold dimension, we can reformulate Eq. (\ref{eq_problem}) as:

\begin{equation}
    \label{eq_opt}
    \begin{split}
        &\min_{\theta, \mathcal{M}} \mathcal{L}(\theta) + \sum_{i=1}^d ||\nabla_{\mathcal{M}} \alpha_i||_{L^2(\mathcal{M})}^2,\\
        & s.t. \  P_{\theta} \subset \mathcal{M}.
    \end{split}
\end{equation}
where
\begin{equation}
    ||\nabla_{\mathcal{M}}\alpha_i||_{L^2(\mathcal{M})} = \left( \int_M ||\nabla_{\mathcal{M}}\alpha_i(\bm{p})||^2d\bm{p} \right)^{1/2}
\end{equation}
To solve this problem, we design an iterative algorithm, named LDM-DN, for optimizing the LDM constrained disentanglement network based on the algorithm for image processing introduced in \cite{LDMM}. Specifically, given $(\theta^k, M^k)$ at step $k$ satisfying $P_{\theta^k}^k \subset \mathcal{M}^k$, step $k+1$ consists of the following sub-steps:
\begin{itemize}
    \item Update $\theta^{k+1}$ and the perturbed coordinate functions $\alpha^{k+1}=(\alpha^{k+1}_1, \cdots, \alpha^{k+1}_d)$ as the minimizers of (\ref{eq_up1}) with the fixed manifold $\mathcal{M}^{k}$:
    \begin{equation}
        \label{eq_up1}
        \begin{aligned}
        &\min_{\theta, \alpha} \mathcal{L}(\theta) + \sum_{i=1}^d ||\nabla_M \alpha_i||_{L^2(M)}^2,\\
        &s.t. \quad  \alpha(P_{\theta^k}) = P_{\theta}
        \end{aligned}
    \end{equation}
    
    \item Update $\mathcal{M}^{k+1}$:
    \begin{equation}
        \label{eq:ignore}
        \mathcal{M}^{k+1} = \{(\alpha_1^{k+1}(\bm{p}), \cdots, \alpha_d^{k+1}(\bm{p})): \bm{p}\in \mathcal{M}^k\}
    \end{equation}
    
    \item Repeat above two sub-steps until convergence.
\end{itemize}

It is noted that if the iteration converges to a fixed point, $\alpha^{k+1}$ will be very close to the coordinate functions, and $\mathcal{M}^{k+1}$ and $\mathcal{M}^k$ will be very close to each other.

Eq. (\ref{eq_up1}) is a constrained linear optimization problem. We can use the alternating direction method of multipliers to simplify the above algorithm as:
\begin{itemize}
    \item Update $\alpha_i^{k+1}, i=1,\cdots,d$, with a fixed $P_{\theta^k}$,
    \begin{equation}
    \label{eq:difficult}
        (\alpha_1^{k+1}, \cdots, \alpha_d^{k+1}) = \arg \min_{\alpha_1,\cdots,\alpha_c \in H^1(\mathcal{M}^n)} \sum_{i=1}^d ||\nabla \alpha_i||^2_{L^2(\mathcal{M}^k)} + \mu ||\alpha(P_{\theta^k}) - P_{\theta^k} + d^k||^2_F.
    \end{equation}
    where 
    \item Update $\theta^{k+1}$,
    \begin{equation}
        \theta^{k+1} = \arg \min_{\theta} \mathcal{L}(\theta) + \mu ||\alpha^{k+1}(P_{\theta^k}) - P_{\theta} + d^k||_F^2
    \end{equation}
    \item Update $d^{k+1}$,
    \begin{equation}
        d^{k+1} = d^k + \alpha^{k+1}(P_{\theta^k}) - P_{\theta^{k+1}}
    \end{equation}
    
\end{itemize}

Using a standard variational approach, the solutions of the objective function (\ref{eq:difficult}) can be obtained by solving the following PDE
\begin{equation}
\label{eq:pde}
    \begin{aligned}
         -\Delta_{\mathcal{M}} u(\bm{p}) + \mu \sum_{y \in \Omega} \delta (\bm{p}-\bm{q})(u(\bm{q})-v(\bm{q})) &= 0, \quad \bm{p}\in \mathcal{M}\\
        \frac{\partial u}{\partial n}(\bm{p}) &= 0, \quad \bm{p} \in \partial \mathcal{M}.
    \end{aligned}
\end{equation}
where $\partial \mathcal{M}$ is the boundary of $\mathcal{M}$, and $n$ is the out normal of $\partial \mathcal{M}$.

Eq. (\ref{eq:pde}) can be solved with the point integral method. For the Laplace-Beltrami equation, the key observation is the following integral approximation:
\begin{equation}
    \int_{\mathcal{M}} \Delta_{\mathcal{M}} u(\bm{q})\bar{R}_t(\bm{p},\bm{q})d\bm{q} \approx -\frac{1}{t} \int_{\mathcal{M}} (u(\bm{p})-u(\bm{q}))R_t(\bm{p},\bm{q})d\bm{q} + 2\int_{\partial \mathcal{M}} \frac{\partial u(\bm{q})}{\partial n} \bar{R}_t(\bm{p},\bm{q})d\tau_{\bm{q}},
\end{equation}
where $t>0$ is a hyper parameter and
\begin{equation}
    R_t(\bm{p},\bm{q}) = C_tR\left(\frac{|\bm{p}-\bm{q}|^2}{4t}\right).
\end{equation}
$R:R^+ \rightarrow R^+$ is a positive $C^2$ function which is integrable over $[0, +\infty)$, and $C_T$ is the normalizing factor
\begin{equation}
    \bar{R}(r) = \int_r^{+\infty} R(s)ds, and \quad \bar{R}_t (\bm{p},\bm{q}) = C_t \bar{R} \left(\frac{|\bm{p}-\bm{q}|^2}{4t} \right)
\end{equation}
We usually set $R(r)=e^{-r}$, then $\bar{R}_t(\bm{p},\bm{q}) = R_t(\bm{p},\bm{q}) = C_t exp\left(\frac{|\bm{p}-\bm{q}|^2}{4t} \right)$ is Gaussian.

Based on the above integral approximation, we approximate the original Laplace-Beltrami equation as:
\begin{equation}
    \int_M(u(\bm{p} - u(\bm{q})) R_t(\bm{p},\bm{q})d\bm{q} + \mu t \sum_{\bm{q}\in \omega} \bar{R}_t(\bm{p},\bm{q})(u(\bm{q})-v(\bm{q})) = 0
\end{equation}
This integral equation is easy to discretize over the point cloud.

To simplify the notation, we denote the patch set $P_{\theta^k} = \{\bm{p}_i\}_{i=1}^m$, where $m$ is the number of patches, in the $k$-th iteration. We assume that the patch set samples the submanifold $\mathcal{M}$ and it is uniformly distributed. Then, the integral equation can be discretized as
\begin{equation}
\label{eq:dis}
    \frac{|\mathcal{M}|}{m} \sum_{j=1}^m R_t(\bm{p}_i, \bm{p}_j)(u_i-u_j) + \mu t \sum_{j=1}^m \bar{R}_t(\bm{p}_i,\bm{p}_j)(u_j-v_j) = 0
\end{equation}
where $v_j = v(\bm{p}_j)$, and $|\mathcal{M}|$ is the volume of the manifold $\mathcal{M}$.

We rewrite Eq. (\ref{eq:dis}) in the matrix form:
\begin{equation}
    (L + \bar{\mu} W)u = \bar{\mu} W v.
\end{equation}
where $v=(v_1,\cdots,v_m)$, $\bar{\mu}=\frac{\mu t m}{|\mathcal{M}|}$, and $L$ is a $m\times m$ matrix,
\begin{equation}
\label{eq_l}
    L=D-W
\end{equation}
$W=(w_{ij}), i,j=1,\cdots,m$ is the weight matrix, $D=diag(d_i)$ with $d_i=\sum_{j=1}^m w_{ij}$, and
\begin{equation}
\label{eq_w}
    w_{ij} = R_t(\bm{p}_i,\bm{p}_j), \quad \bm{p}_i, \bm{p}_j \in P_{\theta^k}, \quad i,j = 1,\cdots,m.
\end{equation}

\begin{algorithm}[!h]
    
	\caption{LDM-DN learning algorithm}
	\algorithmicrequire{$DataSet$ including unpaired training data $\{(x_i^a, y_i)\}_{i=1}^{N_1}$ and/or paired training data $\{(x_j^a, x_j^{gt})\}_{j=1}^{N_2}$, initial network parameters $\theta^0$, initial dual variables $d^0$, hyperparameters $\lambda$ and $\mu$, and the number of training epochs $E$, batch size $bs$.}\\
	\algorithmicensure{ Network parameters $\theta^*$.}
	\begin{algorithmic}[1]
	    \FOR{$e \in \{1,\cdots, E\}$}
	        \FOR{$B \in DataSet$}
	                \STATE Compute the outputs of disentanglement network given a batch of data $B=\{x^a_i, y_i\}_{i=1}^{bs}$, and construct the patch set, $P_{\theta^k}$, as described in Section \ref{sec_patch}.
            		\STATE Compute the weight matrix $W=(w_{ij})$ and $L$ with $P_{\theta^k}$, as in Eqs. (\ref{eq_w}) and (\ref{eq_l}).
            		\STATE Solve the following linear systems to obtain $U$\\
            		$(L+\mu W)U = \mu WV$, where $V=P_{\theta^k}-d^k$.

            		\STATE Update $\theta^{k+1}$ using Adam with the following loss function:\\
            		$\mathcal{J}(\theta) = \mathcal{L}(\theta) + \lambda ||U-P_{\theta^k} + d^k||_F^2$. 
                   \STATE Construct the patch set $P_{\theta^{k+1}}$ with $\theta^{k+1}$ and update $d^{k+1}$ as follows:\\
        		   $\hat{d}^k= d^k + U - P_{\theta^{k+1}}$,\\
        		   $d^{k+1}=(\hat{d}^k-\min(\hat{d}^k))/(\max(\hat{d}^k)-\min(\hat{d}^k))$.\\
        		   \STATE $k \leftarrow k+1$.\\
        	\ENDFOR
    	\ENDFOR
    	\STATE $\theta^* \leftarrow \theta^{(k)}$.
	\end{algorithmic}
	\label{alg:opt}
\end{algorithm}

The final LDM-DN learning algorithm is described in Algorithm \ref{alg:opt}, where we assume that the patch set of all images samples a low-dimensional manifold. However, it is impractical to optimize the LDM problem when the number of patches is very large. To this end, we randomly select a batch of images to construct the patch set, and then estimate the coordinate functions $U$, update the network parameters $\theta$ and dual variables $d$ in each iteration. Thus, in our implementation the number of iterations in training the network is the same as that in the LDM optimization. While practically updating the dual variables $\bm{d}$ in the original LDMM algorithm \cite{LDMM}, the values of $\bm{d}$ usually increase as the number of iterations increases. As the number of iterations is usually very large, the value of the LDM term in step 6 of Algorithm \ref{alg:opt} will become increasingly large, leading to a bad solution. To overcome this problem, the dual variables are normalized in step 7 of Algorithm \ref{alg:opt}.

\subsection{Combination of paired and unpaired learning}
\label{sec_train}

\begin{figure}
    \centering
    \includegraphics[width=0.6\textwidth]{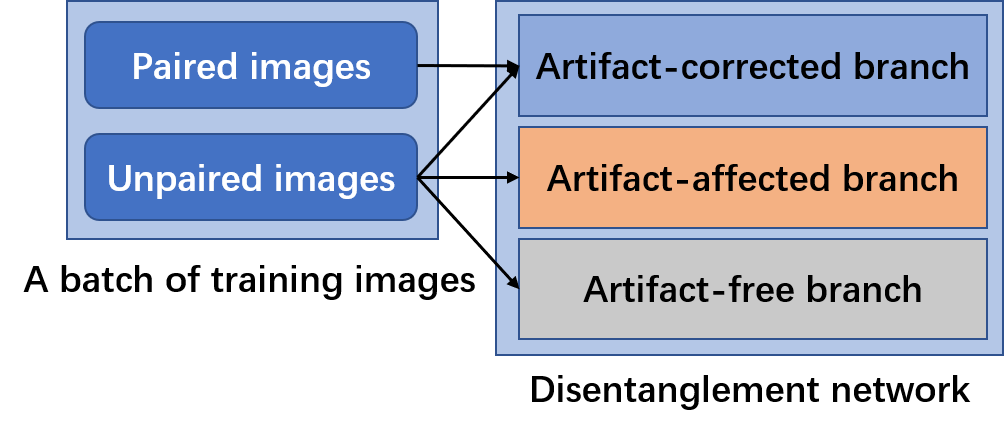}
    \caption{Combination of paired and unpaired learning.}
    \label{fig:train}
\end{figure}

ADN only requires unpaired clinical images for training so that performance degradation of the model  can be avoided when it is first trained on the synthesized dataset and then transferred to a clinical application. However, the GAN loss based unpaired supervision is not strong enough for recovering full image contents details.
On the other hand, although the synthesized data may not perfectly simulate real scenarios, it does provide helpful information via accurate supervision. To benefit from both the paired learning and unpaired learning, here we design a hybrid training scheme. Specifically, during training, both unpaired  clinical images and paired synthetic images are selected to construct a mini-batch, which are then fed to the corresponding branches to optimize the objective functions simultaneously, as shown in Fig. \ref{fig:train}.


\section{Experimental design and results}
\label{sec_experiment}

\subsection{Datasets}
In our experiments we evaluated the proposed method on one synthesized dataset from DeepLesion \cite{yan2018} and one clinical dataset from Spineweb \footnote{spineweb.digitalimaginggroup.ca}, which are the same as those used in ADN \cite{adn}.

For the synthesized dataset, 4,118 artifact-free CT images were randomly selected from DeepLesion. Then, the paired images with and without metal artifacts were synthesized using the method introduced in CNNMAR \cite{zhang2018}. Finally, 3,918 pairs of images were used for training and 200 pairs for testing. For a fair comparison, the images used for training and testing, and all pre-processing processes are the same as those used for ADN \cite{adn}.

For the clinical dataset, 6,170 images with metal artifacts and 21,190 images without metal artifacts are selected for training, and additional 100 images with metal artifacts were selected for evaluation. The criteria for selecting these images are the same as that in the ADN study. Specifically, if an image contains pixels with HU values greater than 2,500 and the number of these pixels is larger than 400, then the image is grouped into the artifact-affected class. The images with the largest HU values less than 2,000 are grouped into the artifact-free class. Furthermore, to study the effectiveness of combining both paired and unpaired supervision, we randomly selected 6,170 images from the artifact-free group. Then, we extracted 6,170 metal objects from the images in the artifact-affected group, and used CatSim \cite{catsim} to simulate the paired images by inserting each extracted metal shape into a selected artifact-free image. Finally, 6,170 synthesized paired images were obtained.

\subsection{Implementation details}

\begin{figure}
    \centering
    \includegraphics[width=0.7\textwidth]{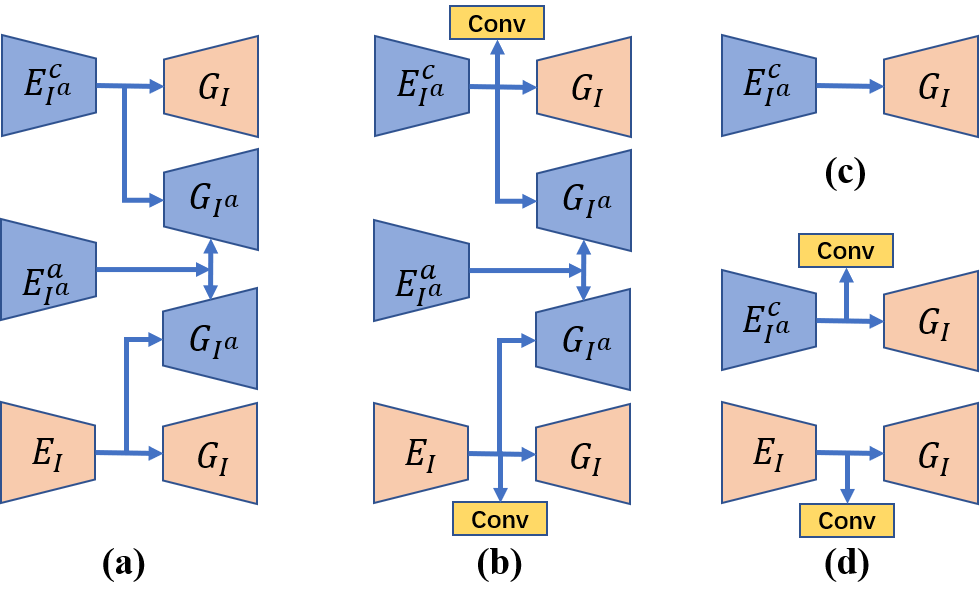}
    \caption{Network architectures of different learning paradigms. $E_{I^a}^c, E_{I^a}^a$ denote the encoders that extract content and artifact features from artifact-affected images. $E_I$ is the encoder that extracts content features from the artifact-free images. $G_I$ and $G_{I^a}$ represent the decoders that output the artifact-free/artifact-corrected and artifact-affected images respectively. The combinations of $E_{I^a}^c \rightarrow G_I$, $E_{I^a}^a \rightarrow G_{I^a}$, and $E_I \rightarrow G_I$ construct artifact-corrected, artifact-affected, and artifact-free branches respectively. $Conv$ denotes a convolutional layer.}
    \label{fig:arcs}
\end{figure}

We implemented different network architecture variants for different learning paradigms, as shown in Fig. \ref{fig:arcs}. For unpaired learning, we use the same architecture as ADN \cite{adn} as shown in Fig. \ref{fig:arcs} (a), and the architectures of all other learning paradigms are the variants of ADN. In Fig. \ref{fig:arcs} (b), to construct the patch set for the LDM constraint, we add two convolutional layers on the top of the encoders in the artifact-corrected and artifact-free branches respectively, as described in Section \ref{sec_patch}. For paired learning only, we simply use the encoder-decoder in the artifact-corrected branch as shown in Fig. \ref{fig:arcs} (c). In combination of paired learning and the LDM constraint, we keep two encoder-decoder branches as shown in Fig. \ref{fig:arcs} (d).

In Fig. \ref{fig:patch} and Fig. \ref{fig:arcs} (b) and (d), the extra convolutional layers are used to compress the channels of the latent code. Specifically, the input image size is $1 \times 256 \times 256$, the down-sampling rate is 8, the matrix of $Z_x$ is of $512 \times 64 \times 64$, the matrix of $Z_x^t$ is of $64 \times 64 \times 64$, the patch size is $8 \times 8$, and the dimension of the point in the patch set is 128.

We implemented the proposed method in PyTorch \footnote{https://pytorch.org/}. To be fair, we keep all hyper parameters the same as those in DAN \cite{adn}. In Algorithm \ref{alg:opt}, we empirically set the batch size $bs=1$, $\mu=0.6$.

\subsection{Results on synthesized dataset}
\label{sec_syn}

\subsubsection{Reimplementation of ADN}
\label{sec_ratio}

\begin{table}
  \renewcommand{\arraystretch}{1.2}
  \renewcommand\tabcolsep{3.5pt}
 \caption{MAR results of the models trained with different ratios of numbers of images in the artifact-free and artifact-affected groups.}
  \centering
  \begin{tabular}{c|ccc}
    \hline
     & ADN-0.85 & ADN-0.50 & ADN-0.15  \\
     \hline
    PSNR & 34.1  & 34.0  & 34.0  \\
    SSIM & 92.8  & 92.8  & 92.9  \\
    \hline
  \end{tabular}
  \label{tab:ratio}
\end{table}

To simulate the unpaired learning, the synthesized paired images were divided into two groups, and then the artifact-affected images were selected from one group and the artifact-free images from the other group. In \cite{adn}, the ratio of numbers of images in these two groups was simply set to 1:1. However, in clinical scenarios, the number of artifact-affected images is much smaller than the number of artifact-free images. Therefore, we evaluated the effectiveness of the ratio to the MAR performance in the unpaired learning setting. In Table \ref{tab:ratio}, \emph{ADN-0.85}, \emph{ADN-0.50} and \emph{ADN-0.15} signify various ratios of artifact-affected images to all images. Table \ref{tab:ratio} shows that there is little difference between the models trained with different ratios between artifact-affected and artifact-free images. In addition, we found that the metrics for the MAR performance would not strictly converge in the unpaired learning setting. Therefore, we selected the best one as the final results, which are better than the reported results in \cite{adn}. In practice, it is also reasonable to select the best performance model on the synthesized images first and then apply the selected model to clinical images. 
As a representative ratio between the artifact-affected and artifact-free images in the clinical conditions, the \emph{ADN-0.15} serves as the baseline in all following experiments.

\begin{table}
  \renewcommand{\arraystretch}{1.2}
  \renewcommand\tabcolsep{3.5pt}
 \caption{Comparison results on DeepLesion.}
  \centering
  \begin{tabular}{c|ccccc|ccccccc}
    \hline
     & \multicolumn{5}{c|}{Paired leaning} & \multicolumn{7}{c}{Unpaired learning}  \\
     & CNNMAE & UNet & cGANMAR  & Sup & \textbf{LDM-Sup}  & CycleGAN & DIP & MUNT & DRIT & ADN & ADN$^*$ & \textbf{LDM-DN} \\
     \hline
    PSNR & 32.5  & 34.8  & 34.1 & 37.6 & \textbf{38.0}  & 30.8 & 26.4 & 14.9 & 25.6 & 33.6 & 34.0 & \textbf{35.0}   \\
    SSIM & 91.4  & 93.1  & 93.4 & 96.1 & \textbf{96.3}  & 72.9 & 75.9 &  7.5 & 79.7 & 92.4 & 92.9 & \textbf{94.2}  \\
    \hline
  \end{tabular}
  \label{tab:compare}
\end{table}

\begin{figure}
    \centering
    \includegraphics[width=0.99\textwidth]{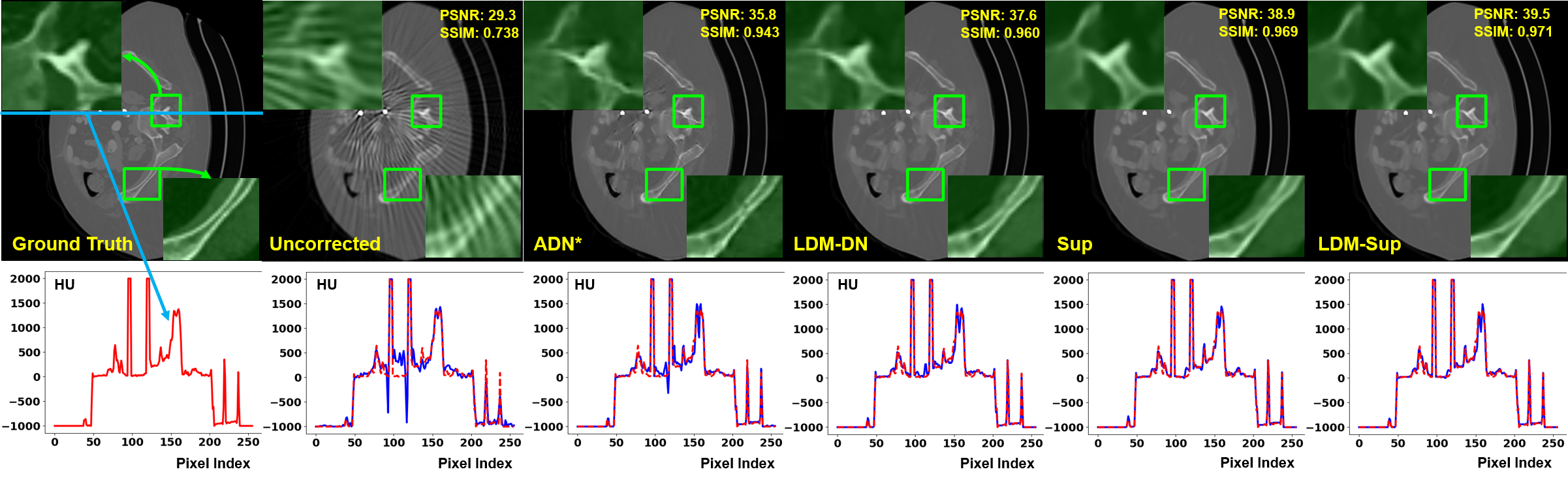}
    \caption{Visual comparison. The metal objects are inserted back by thresholding.}
    \label{fig:syn_results}
\end{figure}

\subsubsection{Comparative results}
On the synthesized dataset, we evaluated the quantitative and qualitative performance of our proposed method as well as the compared methods. For quantitative results, we used the peak signal-to-noise ratio (PSNR) and structural similarity index (SSIM) metrics. Table \ref{tab:compare} gives the comparison results of the proposed method with the competing methods in the paired and unpaired learning settings. In Table \ref{tab:compare}, \emph{ADN}$^*$ is our improved ADN, see Section \ref{sec_ratio} for details. The results show that the proposed \emph{LDM-DN} method outperformed ADN in terms of both PSNR and SSIM metrics in the unpaired learning setting.
For the paired learning part in Table \ref{tab:compare}, \emph{Sup} corresponds to the network architecture in Fig. \ref{fig:arcs} (c), which was trained with the paired data. It is noted that this encoder-decoder architecture contains the skip connections between the encoder and the decoder, which is the same as ADN. \emph{LDM-Sup} adds the LDM constraints to the \emph{Sup} during the paired training, corresponding to the architecture in Fig. \ref{fig:arcs} (d). Although the paired images have accurate pixel-to-pixel supervision, the LDM based learning algorithm can further improve the performance. The above results strongly demonstrate that our proposed LDM-DN algorithm can consistently improve the existing models in the paired and unpaired learning settings.

We also visually compared the results as shown in Fig \ref{fig:syn_results}. 
The visual impressions are consistent with the numerical results. In the unpaired learning setting, although ADN can remove a majority of metal artifacts, the local details were not well preserved. By comparing the results of \emph{ADN*} and \emph{LDM-DN}, an evident improvement was made on these details. Compared with the unpaired learning (\emph{Sup} vs.\emph{ADN*}), the ideal paired learning gave  better results on the synthesized test dataset. In this case, our proposed LDM-DN learning algorithm obtained further improvements, where the boundaries of structures are sharper visually (\emph{LDM-Sup} vs. \emph{Sup}). These results strongly show the effectiveness of the proposed LDM-DN algorithm.

\subsection{Results on clinical dataset}

\begin{figure}
    \centering
    \includegraphics[width=0.99\textwidth]{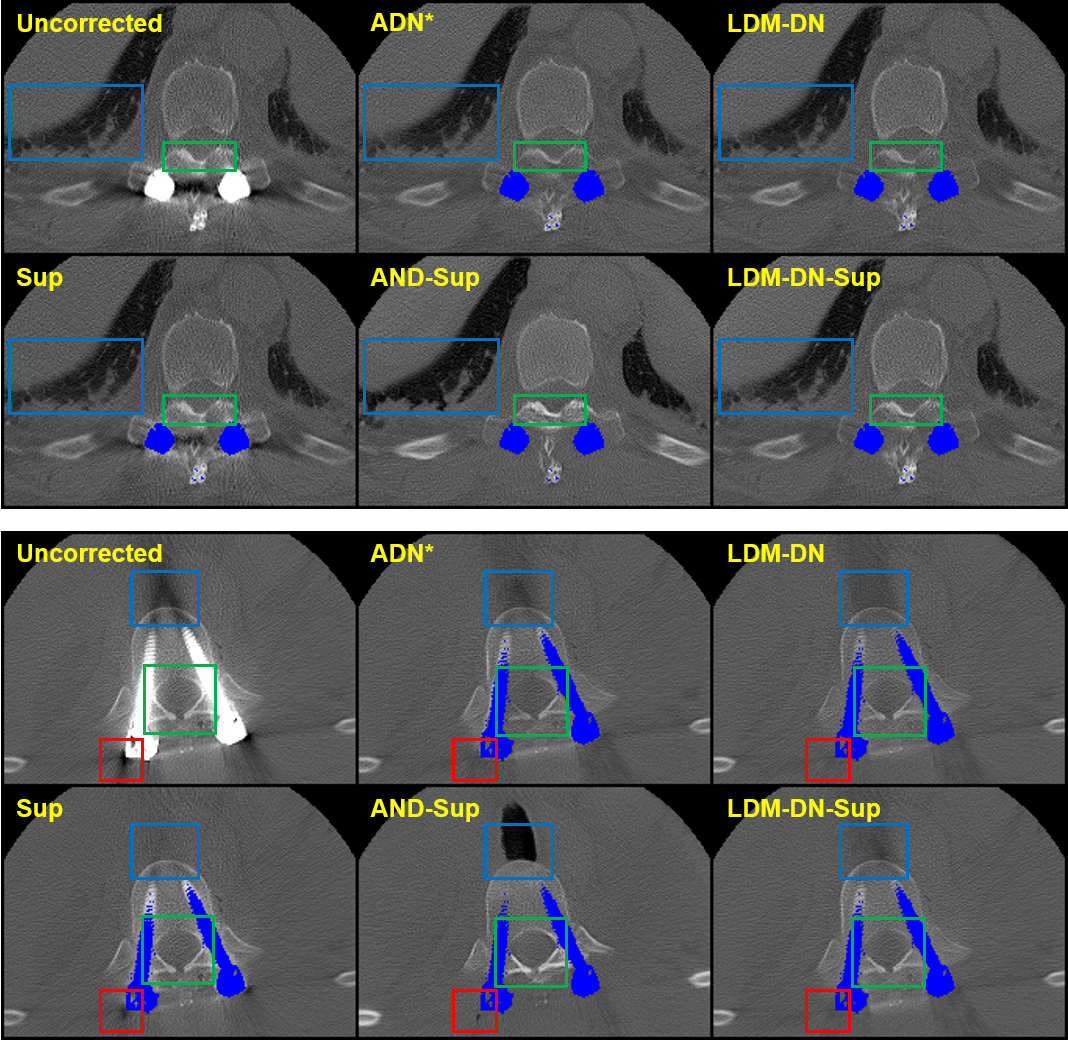}
    \caption{Clinical results from different variants.}
    \label{fig:clinical_results}
\end{figure}

In this subsection, we evaluated the proposed networks on the clinical dataset. As there are not ground truth images for evaluating the performance of the models, here we can only show the visual results of two examples in Fig. \ref{fig:clinical_results}. On the clinical dataset, we see the superiority of the proposed LDM-DN algorithm in preserving (the green boxes for \emph{LDM-DN} vs. \emph{ADN*} in the first example) and recovering (the blue boxes for \emph{LDM-DN} vs. \emph{ADN*} in the second example) local details. We also evaluated the performance of the model trained with synthesized images on the same dataset in a supervised learning manner. As shown in Section \ref{sec_syn}, the results of \emph{ADN*} is better than that in the unpaired learning on synthesized dataset. However, the results in Fig. \ref{fig:clinical_results} shows that the performance of \emph{Sup} is definitely worse than that of the unpaired learning model trained on the clinical dataset, as the synthesized data does not reflect the real conditions. This is consistent with the observation in \cite{adn}. However, although the performance of \emph{Sup} is degraded, it still shows some merits over the unpaired learning methods, such as some structures are sharper (green boxes of \emph{Sup} vs. \emph{ADN*} and \emph{LDM-DN} for both examples) and some regions are better (blue boxes of \emph{Sup} vs. \emph{ADN*} and \emph{LDM-DN} in the second example). 
Combining \emph{ADN} and \emph{Sup} leads to an over-correction as shown in Fig. \ref{fig:clinical_results}, where some structures are over sharper (green boxes of \emph{ADN-Sup} vs. others for both examples) and some regions are over dark (blue boxes of \emph{ADN-Sup} vs. others for both examples). We attribute these results to that \emph{ADN} help reduce the under-correction of \emph{Sup} in some regions (red boxes of \emph{ADN-Sup} vs. \emph{Sup} for the second example) but simultaneously resulting in over-correcting some other regions (blue boxes of \emph{ADN-Sup} vs. \emph{Sup} for the second example).
Therefore, we propose to combine all merits of unpaired learning, paired learning and LDM through a hybrid learning scheme. As the results of \emph{LDM-DN-Sup} shown in Fig. \ref{fig:clinical_results}, it can inherit all good points as analyzed above. Particularly, compared with \emph{ADN-Sup}, LDM in \emph{LDM-DN-Sup} constrains that the structurally similar patches, especially the adjacent patches, to be coherent without dramatic changes to be too dark or too bright. The above results on the clinical dataset strongly demonstrate the effectiveness of LDM and the superiority of the hybrid training scheme.

\section{Conclusion}
\label{sec_conclusion}

We have proposed an LDM constrained disentanglement network for MAR. Specifically, we have designed a LDM-DN learning algorithm to simultaneously optimize the objective functions of deep neural networks and constrain the recovered images to have a low-dimensional patch manifold representation. The LDM-DN algorithm can effectively help preserve and recover structural details in CT images. Moreover, we have investigated both paired and unpaired learning based models for MAR, showing their relative advantages. Finally, we have designed a hybrid optimization scheme to combine paired learning, unpaired learning and LDM-DN learning algorithm for integrating their advantages. The experimental results on synthesized and clinical datasets have strongly demonstrated the superiority of the proposed method. We believe that the proposed LDM-DN algorithm has a great potential to solve various CT MAR problems.

\bibliographystyle{unsrt}
\bibliography{references}  

\end{document}